\documentclass{PoS}
\usepackage{subfigure}
\usepackage{amsmath,amssymb}
\usepackage{enumerate,enumitem}
\usepackage{here}

\title{Towards higher order numerical stochastic perturbation computation applied to the twisted Eguchi-Kawai model
\footnote{This work is partly supported by Priority Issue 9 to be tackled by using Post K Computer. Numerical simulations are performed on the ITO supercomputer system of Kyushu university.}}

\ShortTitle{Towards higher order NSPT computation applied to the TEK model}

\author{
Antonio Gonz\'{a}lez-Arroyo$^{a,b}$, Issaku Kanamori$^{c}$, \speaker{Ken-Ichi Ishikawa}$^{d,e}$, Kanata~Miyahana$^{d}$, Masanori~Okawa$^{d}$, Ryoichiro~Ueno$^{d}$\\
$^a$~Instituto de F\'{i}sica Te\'{o}rica UAM/CSIC, Nicol\'{a}s Cabrera 13-15, E-28049 Universidad Aut\'{o}noma de Madrid, Madrid, Spain\\
$^b$~Departamento de F\'{i}sica Te\'{o}rica, M\'{ó}dulo 15, Cantoblanco,         E-28049 Universidad Aut\'{o}noma de Madrid, Madrid, Spain\\
$^c$~RIKEN Center for Computational Science, Kobe 650-0047, Japan\\
$^d$~Graduate School of Science, Hiroshima University, Higashi-Hiroshima 739-8526, Japan\\
$^e$~Core of Research for the Energetic Universe, Hiroshima University, Higashi-Hiroshima 739-8526, Japan\\
        E-mail: \email{ishikawa@theo.phys.sci.hiroshima-u.ac.jp}}

\abstract{
 {\normalsize        %
 \vspace*{-38.0em}   %
 \begin{flushright}  
   HUPD-2001, IFT-UAM/CSIC-20-5
 \end{flushright}    %
 \vspace*{35.0em}\     %
 }                   %
We have evaluated perturbation coefficients of Wilson loops up to $O(g^8)$ for the four-dimensional
twisted Eguchi-Kawai model using the numerical stochastic perturbation theory (NSPT) in~\cite{Gonzalez-Arroyo:2019zfm}.
In this talk we present a progress report on the higher order calculation up to $O(g^{63})$,
for which we apply a fast Fourier transformation (FFT) based convolution algorithm to the multiplication
of polynomial matrices in the NSPT aiming for higher order calculation. 
We compare two implementations with the CPU-only version and the GPU version of the FFT based convolution algorithm, and 
find a factor 9 improvement on the computational speed of the NSPT algorithm with SU($N=225$) at $O(g^{31})$.
The perturbation order dependence of the computational time, we investigate it up to $O(g^{63})$, 
shows a mild scaling behavior on the truncation order.}

\FullConference{37th International Symposium on Lattice Field Theory - Lattice2019\\
		16-22 June 2019\\
		Wuhan, China}

\begin{document}

\section{Introduction}
The numerical stochastic perturbation theory (NSPT) is one of the numerical method to compute 
a higher order perturbative series of lattice quantum field theories~\cite{DiRenzo:1993hs,DiRenzo:1994av,DiRenzo:1994sy,DiRenzo:1995qc,Burgio:1997hc,DiRenzo:2004hhl}.
Investigating the higher order behavior of perturbation coefficients in terms  of expansion order 
could reveal the non-perturbative properties such as instanton, non-trivial complex saddles, and renormalon effects.

The large $N$ limit of Yang-Mills theory is one of the tools to investigate the non-perturbative aspects 
keeping the non-perturbative properties with a great simplification~(see for recent studies~\cite{Lat19Plenary}).
To tackle the non-perturbative aspects of the large-$N$ Yang-Mills theory through the higher order behavior 
of the perturbative coefficients, 
we have evaluated the coefficients of Wilson loops up to $O(g^8)$ for the four-dimensional twisted 
Eguchi-Kawai (TEK) model~\cite{GonzalezArroyo:1982ub,GonzalezArroyo:1982hz} using the NSPT~\cite{Gonzalez-Arroyo:2019zfm}.
The eighth order of the expansion is not yet sufficient to study the higher order behavior, though, 
we found that the variance of the coefficients is well controlled by the factorization property of large-$N$ expansion,
and there is no \textit{Pepe} effect by which we can access the higher order coefficients with a moderate statistics.
We also investigated the computational cost of the NSTP for the TEK model in the large-$N$ limit, and found that
the computational cost of the perturbation series multiplication can be an obstacle to access the higher order coefficients in the large-$N$ limit.
This is because that the cost scales as $N_{\mathrm{trunc}}^2$ for the multiplication of two perturbative series 
truncated at $O(g^{N_{\mathrm{trunc}}})$ with the naive multiplication algorithm.

In order to improve the scaling property in a higher order perturbation truncation, we implement a Fast Fourier Transformation based 
convolution algorithm (FFT-CONV) for the perturbation polynomial multiplication.
Because we expand the $SU(N)$ matrix perturbatively in the TEK model, the series expansion becomes polynomial matrices,
 i.e. the coefficients are $N\times N$ matrices, and the implementation of the FFT-CONV is not trivial.

In this talk we present the FFT based convolution algorithm for the polynomial matrix multiplication, and apply this algorithm to the NSPT algorithm of the TEK model.
We first compare the naive-multiplication algorithm and the FFT-CONV algorithm on a CPU architecture machine. 
Having established a high efficiency of the FFT-CONV algorithm, we further implement the FFT-CONV for a GPU system, 
and compare the performance between the CPU and GPU versions.

\section{FFT based convolution algorithm}
The TEK model is defined by
\begin{align}
  Z&=\int \prod_{\mu=1}^{4}dU_{\mu} \exp\left[-\beta\sum_{\mu\ne\nu}\mathrm{Tr}\left[1-z_{\mu\nu}U_\mu U_\nu U_\mu^{\dag} U_{\nu}^{\dag}\right]\right],
\end{align}
where $U_\mu$ are SU($N$) matrices and $z_{\mu\nu}$ is an anti-symmetric tensor called the twist-tensor~\cite{GonzalezArroyo:1982ub,GonzalezArroyo:1982hz}. 
The non-perturbative simulation algorithm based on the molecular dynamics algorithm, such as HMC, 
relies on the following equation of motion.
\begin{alignat}{2}
\dot{U}_{\mu} &= i P_{\mu} U_{\mu}, &\qquad
F_\mu  &= i\beta\left[ S_{\mu}-S_{\mu}^{\dag}-\mathrm{Tr}[S_{\mu}-S_{\mu}^{\dag}]/N\right],
\notag\\
\dot{P}_\mu &= F_{\mu},&
S_\mu &= U_\mu\left[ \sum_{\nu\ne\mu}\left(z_{\mu\nu}U_{\nu}U_{\mu}^{\dag}U_{\nu}^{\dag}-z_{\mu\nu}^*U_{\nu}^{\dag}U_{\mu}^{\dag}U_{\nu}\right)\right],
\label{eq:EOM}
\end{alignat}
where dot $\dot{\ \ }$ means the fictitious time derivative.
The NSPT algorithm is derived from the equation of motion by expanding them perturbatively in $g=1/\sqrt{\beta}$.
Substituting $U_{\mu}=\sum_{k=0}g^k U_{\mu}^{(k)}$ and $P_{\mu}=(1/g)\sum_{k=1}g^k P_{\mu}^{(k)}$ into 
\eqref{eq:EOM}, expanding them and rescaling the fictitious time, we have
\begin{align}
  \dot{U}^{(k)}_{\mu} &= i (P_{\mu} \star U_{\mu})^{(k)}, \quad \dot{P}^{(k)}_\mu = F^{(k)}_{\mu}, \quad \mbox{[$+$ stochastic gauge fixing part]}
\notag\\
F^{(k)}_\mu &= i\beta\left[ S^{(k)}_{\mu}-{S^{(k)}_{\mu}}^{\dag}-\mathrm{Tr}[S^{(k)}_{\mu}-{S^{(k)}_{\mu}}^{\dag}]/N\right],
\notag
\\
S^{(k)}_\mu &= \left(U_\mu\star
\left[ \sum_{\nu\ne\mu}\left(z_{\mu\nu}  U_{\nu}\star U_{\mu}^{\dag}\star U_{\nu}^{\dag}
                            -z_{\mu\nu}^*U_{\nu}^{\dag}\star U_{\mu}^{\dag}\star U_{\nu}\right)\right]\right)^{(k)},
\end{align}
where the details of the stochastic gauge fixing part are omitted.
The tower of the equations is truncated at $O(g^{N_{\mathrm{trunc}}})$
and is solved with a symplectic integrator with periodic momentum refreshment to yield 
the statistical ensemble for the perturbative coefficients. 
The star $\star$ product is the convolution product explained in the following.

Let $A,B,C$ are $N\times N$ polynomial matrices truncated at $O(g^{N_{\mathrm{trunc}}})$ defined by
\begin{align}
  A &= \sum_{k=0}^{N_{\mathrm{trunc}}} g^k A^{(k)}, \quad
  B  = \sum_{k=0}^{N_{\mathrm{trunc}}} g^k B^{(k)}, \quad
  C  = \sum_{k=0}^{N_{\mathrm{trunc}}} g^k C^{(k)}.
\end{align}
The matrix product $C=AB$ is expanded and $C^{(k)}$'s are expressed with the convolution of $A^{(k)}$ and $B^{(k)}$ as
\begin{align}
  C^{(k)} &= \sum_{\ell=0}^{k} A^{(\ell)}B^{(k-l)} \equiv (A\star B)^{(k)},
\end{align}
where we define $\star$ as the convolution product.  
As the coefficients $A^{(k)}$ and $B^{(k)}$ are $N\times N$ matrices the ordering of multiplication have to be kept.
The computational cost of the naive convolution scales with  $N^3 \times N_{\mathrm{trunc}}^2$, 
where $N^3$ comes from the $N\times N$ matrix-matrix multiplication and $N_{\mathrm{trunc}}^2$ from the naive convolution.

We introduce the FFT based convolution algorithm (FFT-CONV) in the following.
In order to treat the truncated polynomial multiplication, the FFT-CONV requires twice a longer series padded 
with zero coefficients to avoid the overlapping effect due to the periodicity of FFT. 
This type of method is called the linear convolution algorithm.
The FFT is known to be very efficient when the length of data is a power of 2.
We restrict the order of truncation $N_{\mathrm{trunc}}$ to be $N_{\mathrm{trunc}} = N_p-1$ with $N_p=2^s$ and $s\in \mathbb{N}$.
The FFT-CONV for the polynomial matrix multiplication is summarized as follows.
\begin{enumerate}[leftmargin=4em]
\item[\bf Step 1] Double the perturbative series length with zero coefficient matrices.
    \begin{align}
\left\{A^{(0)},A^{(1)},\dots,A^{(N_p-1)}\right\} \Rightarrow
\left\{A^{(0)},A^{(1)},\dots,A^{(N_p-1)}, A^{(N_p)},\dots, A^{(2N_p-1)}\right\},
    \end{align}
with $A^{(j)}=0, (j=N_p,\dots,2 N_p -1)$. Double the length for $B^{(k)}$ similarly.
\item[\bf Step 2] Do FFT with the length $2N_p$ on $\left\{A^{(k)}\right\}$ 
      and $\left\{B^{(k)}\right\}$ to obtain 
the Fourier coefficients $\left\{\tilde{A}^{(p)}\right\}$ and $\left\{\tilde{B}^{(p)}\right\}$.
\begin{align}
  \tilde{A}^{(p)} &= \sum_{k=0}^{2N_p-1} A^{(k)} w^{kp},\quad
  \tilde{B}^{(p)}  = \sum_{k=0}^{2N_p-1} B^{(k)} w^{kp},\quad
w \equiv  \exp\left(\dfrac{2\pi i}{2N_p}\right).
\end{align}
\item[\bf Step 3] Evaluate $\left\{\tilde{C}^{(p)}\right\}$ with the Fourier-mode-wise multiplication.
        \begin{align}
          \tilde{C}^{(p)}=\tilde{A}^{(p)}\tilde{B}^{(p)},
        \end{align}
\item[\bf Step 4] Do inverse FFT on $\left\{\tilde{C}^{(p)}\right\}$ to obtain $\left\{ C^{(k)}=(A\star B)^{(k)}\right\}$.
          \begin{align}
            C^{(k)} = \dfrac{1}{2N_p}\sum_{p=0}^{2N_p-1} \tilde{C}^{(p)} w^{-kp}.
          \end{align}
\item[\bf Step 5] Discard higher order coefficients $C^{(k)}, (k=N_p,\dots,2N_p-1)$.
\end{enumerate}
The computational cost of the FFT-CONV algorithm scales with $N^3 \times N_{\mathrm{trunc}} \log(N_{\mathrm{trunc}})$.

Several tricks should be inserted in the above steps to get a more better numerical performance in larger $N$.
Most of the FFT libraries, such as FFTW, would be optimized for transforming scalar data. 
It is the case we have to move the perturbation order index to the major index of the array to extract the best performance of the FFT library.
Similarly the matrix-matrix multiplication scans the gauge group indices, the group indices should be located in the major indices
of the array to have the best performance.
Therefore we need to permute the order of the gauge group and perturbation order indices before and after 
the FFT, mode-wise matrix-matrix multiplication, and inverse FFT steps to maximize the performance of fundamental operations of FFT and matrix-matrix multiplication.

A further improvement can be achieved by noticing the independence or parallelism of the operations.
The FFT and inverse FFT operations can be carried out for each gauge group matrix element independently. 
The batch mode of FFT can be utilized if the library supports. 
Similarly the matrix-matrix multiplication for each frequency mode is also independent each other,
the batch mode of the matrix-matrix multiplication is also applicable if it supports.

We implemented the FFT-CONV algorithm for both CPU and GPU architectures. 
As the benchmark we measure and compare the timing of NSPT runs for 10 trajectories 
at several $N$ for SU($N$) and truncation order $N_{\mathrm{trunc}}$.
The machines and libraries, and the benchmark results are shown in the next section.

\begin{table}[t]
    \centering
{\scriptsize
    \begin{tabular}{c|c|c}\hline
           & ITO Subsystem A & ITO Subsystem B \\ \hline\hline
  CPU/node &  
             \begin{tabular}{c}
              Intel Xeon Gold 6154 (Skylake-SP) \\
              (3.0 GHz (Turbo 3.7 GHz), 18 core)$\times$2
             \end{tabular} &
             \begin{tabular}{c}
              Intel Xeon Gold 6140 (Skylake-SP)\\
              (2.3 GHz (Turbo 3.7 GHz), 18 core)$\times$2
             \end{tabular} \\\hline
  GPU/node & NA &
             \begin{tabular}{c}
             NVIDIA Tesla P100 (Pascal) \\
             (1,328 - 1,480 MHz, 3584 CUDA cores)$\times$4 
             \end{tabular} \\\hline
\begin{tabular}{l}
 DP peak\\perf./node 
\end{tabular} & 3,456 GFLOPS  &
             \begin{tabular}{c}
                 CPU: 2,649.6 GFLOPS\\
                 GPU: 5.3 TFLOPS $\times$ 4 = 21.2 TFLOPS
             \end{tabular} \\\hline
\begin{tabular}{l}
 Memory\\B.W./node 
\end{tabular} & 255.9 GB/sec  &
             \begin{tabular}{c}
                 CPU: 255.9 GB/sec\\
                 GPU: 732GB/sec $\times$ 4 = 2928 GB/sec
             \end{tabular} \\
\hline
    \end{tabular}
}
    \caption{System description of the ITO supercomputer.}
    \label{tab:machine}
\end{table}

\section{Numerical Tests}

\subsection{Machine and Libraries}

We employ two subsystems A and B of the ITO supercomputer system at Kyushu University for the benchmark. 
The details of the systems are tabulated in Table~\ref{tab:machine}. The subsystem A is used for benchmarking 
the CPU version and the subsystem B for the GPU version. We employ single node.
In order to utilize the four GPU cards on the node, 
we assign computational tasks, which are independent in the four-dimensional direction, on each GPU card so that
the FFT-CONV algorithm for a matrix-matrix multiplication is carried out within a GPU card.

\begin{table}[t]
    \centering
{\scriptsize
    \begin{tabular}{c|c|c}\hline
&
\begin{tabular}{c}
   CPU version\\(Subsystem A)
\end{tabular} &
\begin{tabular}{c}
   GPU version\\(Subsystem B)
\end{tabular} \\ \hline\hline
 Compiler  & 
Intel Fortran v18.0.0.128  &
            \begin{tabular}{c}
             Intel Fortran v18.0.0.128\\
             + CUDA v9.1.85 
            \end{tabular}  \\\hline
 FFT       &\begin{tabular}{c}
             Intel MKL FFT (Dfti)\\
             batch mode available
            \end{tabular} &
            \begin{tabular}{c}
             cuFFT\\
             batch mode available
            \end{tabular} \\\hline
 ZGEMM     &\begin{tabular}{c}
             Intel MKL BLAS\\
             \texttt{ZGEMM\_BATCH}\\
             batch mode
            \end{tabular} &
             \begin{tabular}{c}
             cuBLAS\\
             \texttt{cublasZgemm}\\
             batch mode using CUDA streams
            \end{tabular} \\\hline
 Transpose  &\begin{tabular}{c}
             Intel MKL BLAS\\
             \texttt{MKL\_ZOMATCOPY}\\
             BLAS extension
            \end{tabular} &
             \begin{tabular}{c}
             cuBLAS\\
             \texttt{cublasZgeam}\\
             BLAS extension
            \end{tabular}\\\hline
    \end{tabular}
}
    \caption{Compilers and libraries}
    \label{tab:libs}
\end{table}

Table~\ref{tab:libs} shows the compilers and libraries for FFT and matrix-matrix multiplication.
The ZGEMM subroutine from the BLAS library is employed for the matrix-matrix multiplication.
In order to permute the group indices and the perturbation index in the FFT-CONV algorithm, 
we employ BLAS extension subroutines to extract the best performance on each architecture.

\begin{table}[t]
    \centering
{\small
    \begin{tabular}{ccll}\hline
                & MD steps &  $N$ of SU($N$) & $N_{\mathrm{trunc}}$ \\ \hline\hline
CPU(Naive-CONV) & 40  & 121, 169, 225      & 7, 15, 31 \\
CPU(FFT-CONV)   & 40  & 121, 169, 225      & 7, 15, 31, 63 \\
GPU(FFT-CONV)   & 40  & 121, 169, 225, 289 &    15, 31, 63 \\\hline
    \end{tabular}
}
    \caption{Parameters}
    \label{tab:params}
\end{table}

\subsection{Benchmark Results}
The parameters used in benchmarking are shown Table~\ref{tab:params}.
All CPU cores are utilized using OpenMP multithreading.

We first measure the performance of the naive convolution (Naive-CONV) algorithm and 
the FFT-CONV algorithm on the CPU system.
Figure~\ref{fig:naivevsfft} shows the timing for 10 trajectories of the NSPT algorithm at $N_{\mathrm{trunc}}=31$
as a function of the color size $N$.
Total time (black squares) accumulates the breakup of various functions involved in the NSPT algorithm. 
The most timing consuming part is the gauge momentum update (green up-triangles) 
and the gauge link update (cyan down-triangles). The FFT-CONV version (b) is faster by a factor 7 
than the Naive-CONV version (a) at $N=225$ on the CPU architecture.
The gray dashed line shows the scaling with $N^3$, which originates from the computational cost of matrix-matrix multiplication.
The total timing of the FFT-CONV fits with the scaling.

Having observed a high performance for the FFT-CONV on the CPU, 
we compare the FFT-CONV between the CPU version (Fig.~\ref{fig:naivevsfft} (b)) and the GPU version (Fig.~\ref{fig:fftgpu} (a)).
A factor 9 improvement on the timing is achieved at $N=225, N_{\mathrm{trunc}}=31$ by employing the GPU cards. 
The scaling is consistent with $N^3$ (gray dashed line).
Figure~\ref{fig:fftgpu} (b) shows the breakup on the fundamental operations of FFT, GEMM, and transpose, involved in the NSPT.
The most timing consuming part is in the matrix-matrix multiplication of \texttt{ZGEMM}. 
On the GPU architecture, \texttt{ZGEMM} is quite highly optimized in the cuBLAS library, the timing is already at the best performance.

Figure~\ref{fig:fftvsNtrunc} shows the timing as a function of truncation order $N_{\mathrm{trunc}}$.
The GPU version shows a milder scaling behavior, while the CPU version has less trivial behavior in the lower truncation region.
Looking at the timing breakup with fundamental operations, the CPU version has a large overhead in the transpose operation, 
especially in the lower truncation order. 
The shape of the array with the color and perturbation order indices is quite narrow at lower $N_{\mathrm{trunc}}$,
the CPU version might have a penalty in the transpose operation for such a narrow array because its performance is sensitive 
to the cache and  the memory subsystem of the architecture. 
On the other hand the GPU version does not show such a irregular behavior, we expect that 
the transpose operation \texttt{cublasZgeam} of cuBLAS extension is well optimized even for such narrow arrays.

The peak performance ratio of the subsystem B (GPU) to A (CPU) is about 6, and the peak memory bandwidth ratio (B/A) is about 11.
Observed improvement factor of 9 at $N=225, N_{\mathrm{trunc}}=31$ is roughly at the arithmetic or geometric average of the two peak ratio.
We observed a consistent performance improvement using the GPU architecture for the FFT-CONV algorithm.

\newcommand{\figscale}{0.55}
\begin{figure}[t]
    \centering
\vspace*{-1em}
\subfigure[Naive-CONV (CPU), function breakup]{\includegraphics[scale=\figscale,clip,trim=-10 5 -10 0]{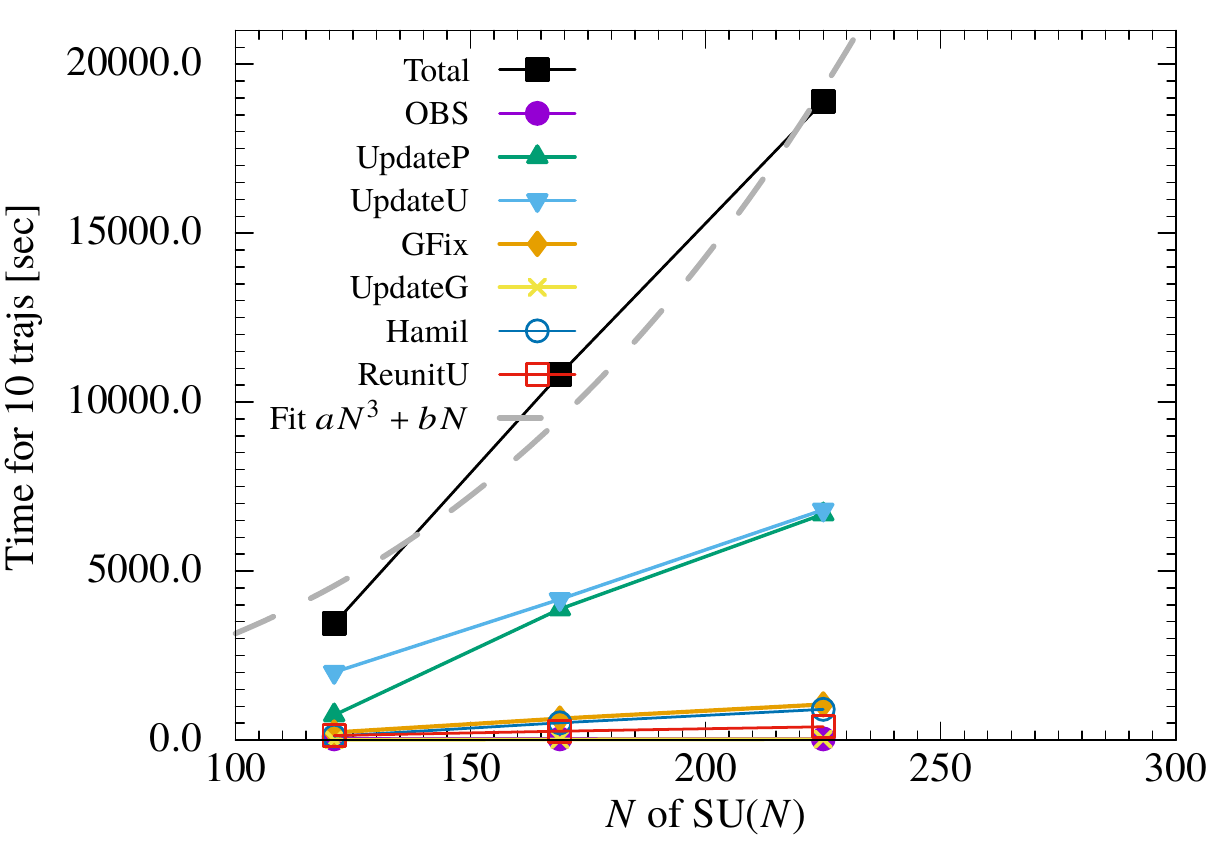}}
\subfigure[FFT-CONV (CPU), function breakup]{  \includegraphics[scale=\figscale,clip,trim=-10 5 -10 0]{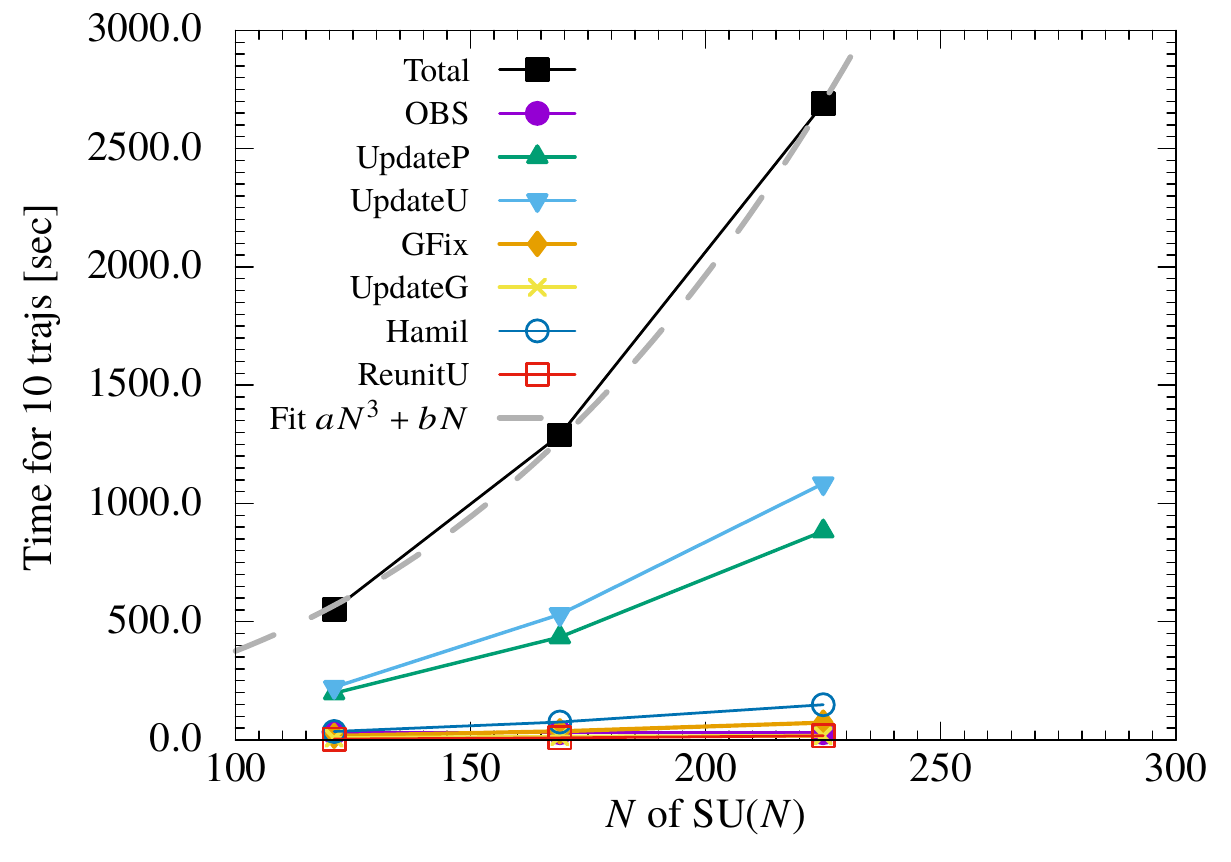}}
\vspace*{-.5em}
    \caption{Timing comparison between the Naive-CONV and FFT-CONV on the CPU at $N_{\mathrm{trunc}}=31$.}
    \label{fig:naivevsfft}
\end{figure}
\begin{figure}[t]
    \centering
\subfigure[FFT-CONV (GPU), function breakup]{             \includegraphics[scale=\figscale,clip,trim=-10 5 -10 0]{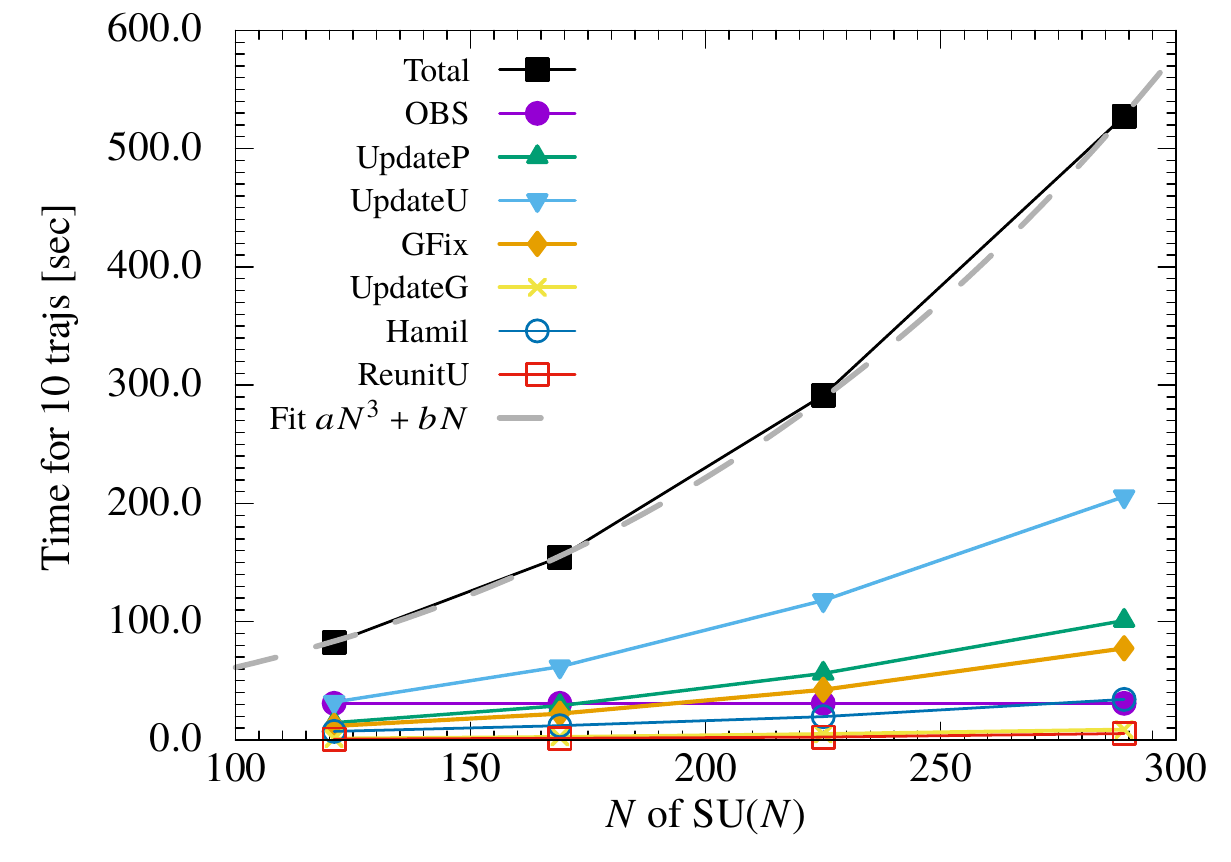}}
\subfigure[FFT-CONV (GPU), fundamental operation breakup]{\includegraphics[scale=\figscale,clip,trim=-10 5 -10 0]{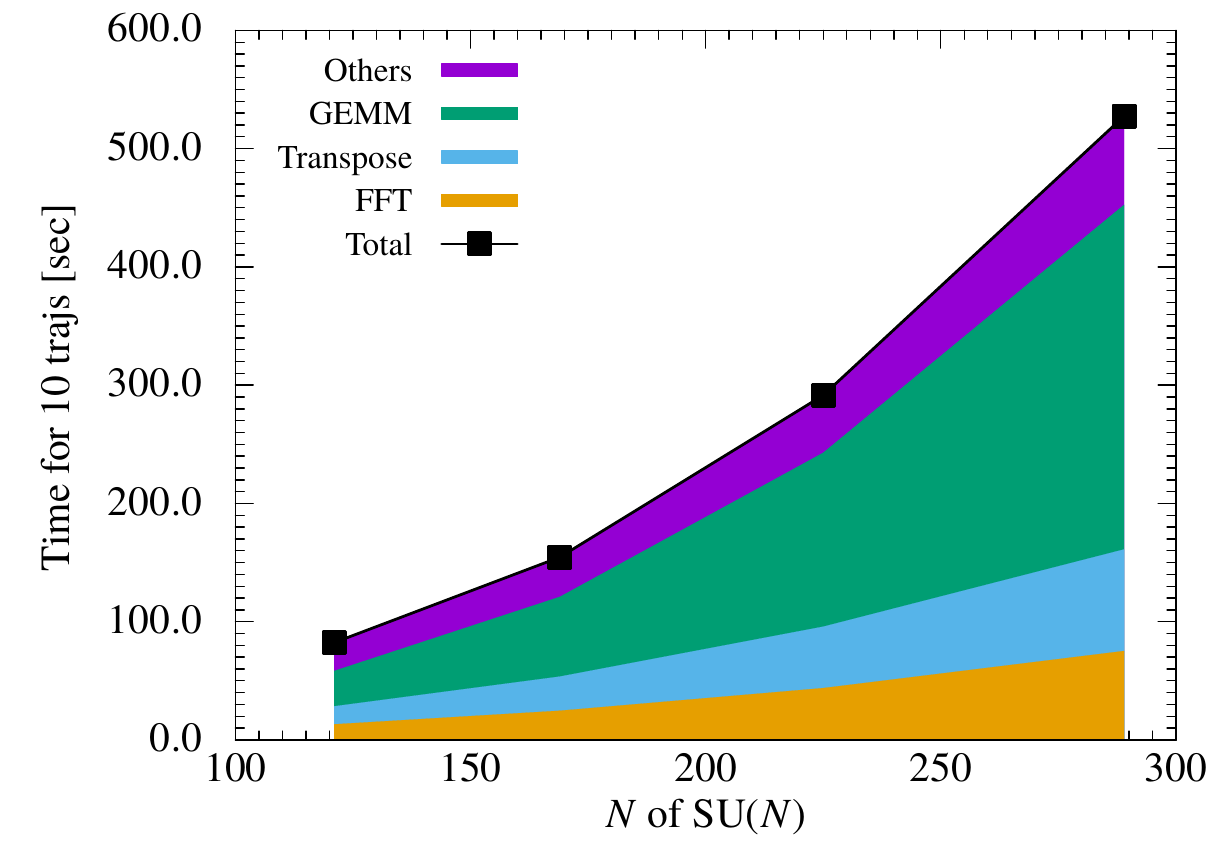}}
\vspace*{-.5em}
    \caption{Timing of the FFT-CONV on the GPU architecture at $N_{\mathrm{trunc}}=31$.}
    \label{fig:fftgpu}
\end{figure}

\begin{figure}[t]
    \centering
\vspace*{-1em}
\subfigure[FFT-CONV (GPU)]{\includegraphics[scale=\figscale,clip,trim=-10 5 -10 0]{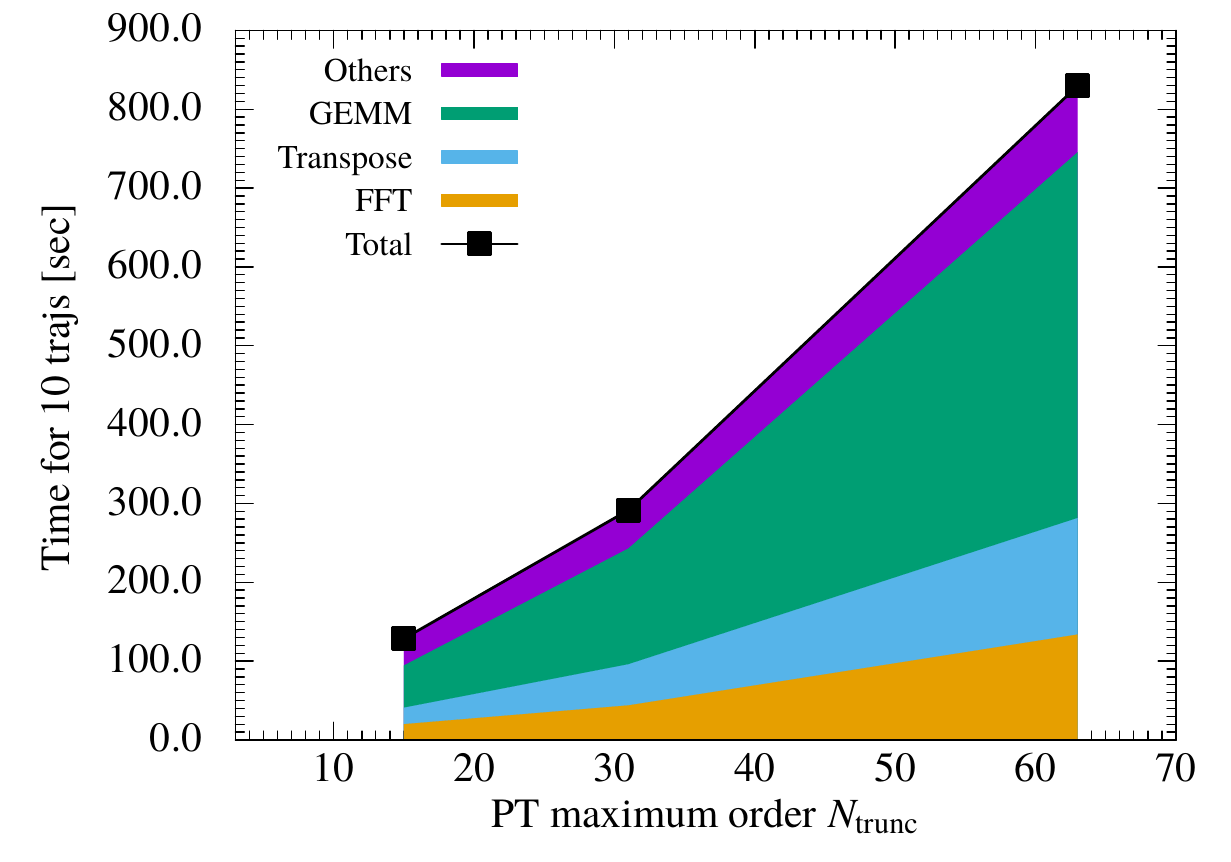}}
\subfigure[FFT-CONV (CPU)]{\includegraphics[scale=\figscale,clip,trim=-10 5 -10 0]{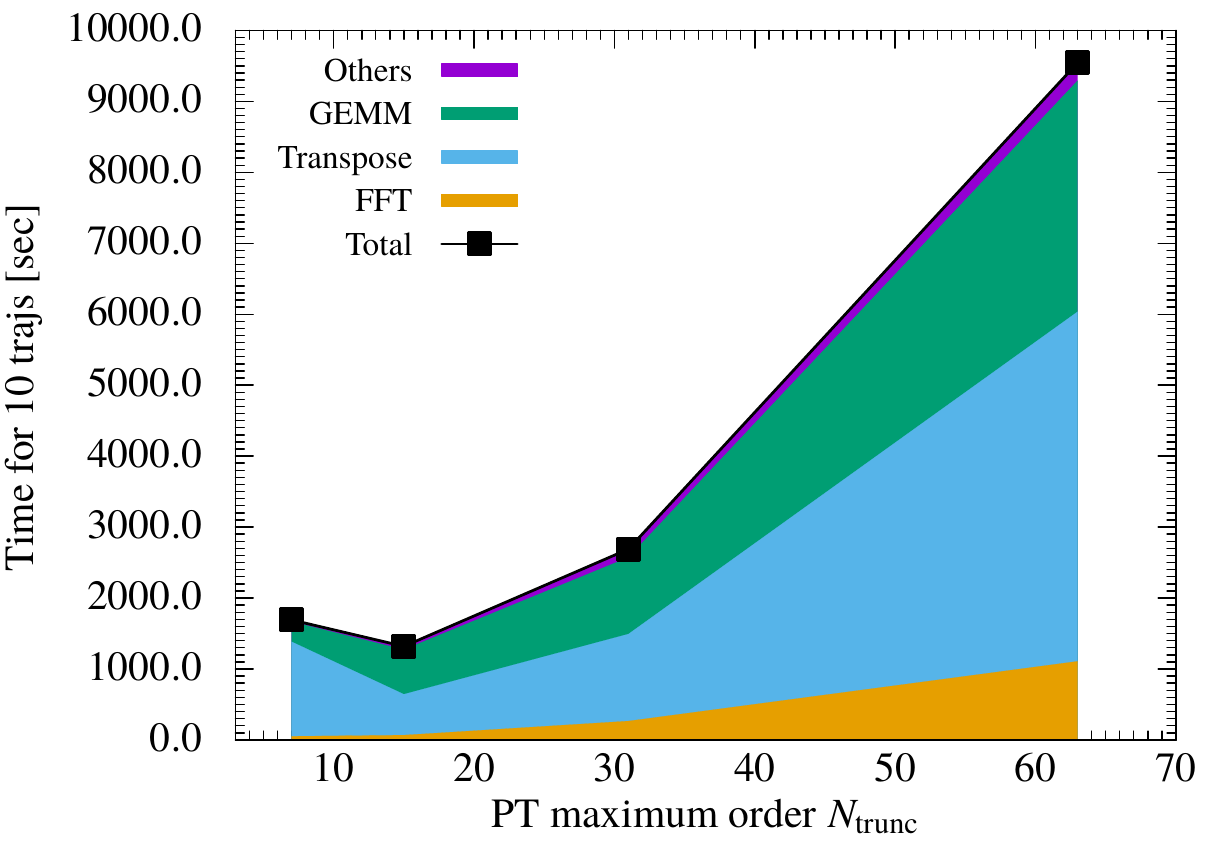}}
\vspace*{-.5em}
    \caption{Timing comparison of the FFT-CONV between GPU version and CPU version at $N=225$.}
    \label{fig:fftvsNtrunc}
\end{figure}

\section{Summary}
We have implemented the FFT based convolution algorithm for the polynomial matrix multiplication in order to improve the performance of the NSPT algorithm 
for the TEK model in the large $N$ region. The FFT-CONV performs quite well even on the CPU system, we can further accelerate it by employing GPU cards.
The TEK model is a reduced model and has no lattice sites to be parallelized, GPU systems is quite preferable to the TEK model as we observed in this benchmark report.

\end{document}